\documentclass[aps,preprint,nofootinbib]{revtex4}
\usepackage{epsfig,hyperref}
\usepackage{graphics}
\usepackage{graphicx}
\usepackage[english]{babel}
\usepackage{amssymb}
\usepackage{amsmath}
\usepackage{subfigure}

\usepackage[normalem]{ulem}

\newcommand{\newc}{\newcommand}
\newc{\beq}{\begin{equation}}
\newc{\eeq}{\end{equation}}
\newc{\barray}{\begin{eqnarray}}
\newc{\earray}{\end{eqnarray}}
\newc{\barrayn}{\begin{eqnarray*}}
\newc{\earrayn}{\end{eqnarray*}}
\newc{\bcenter}{\begin{center}}
\newc{\ecenter}{\end{center}}
\newc{\ket}[1]{| #1 \rangle}
\newc{\bra}[1]{\langle #1 |}
\newc{\0}{\ket{0}}
\newc{\mc}{\mathcal}
\newc{\er}[1]{(\ref{eq:#1})}
\newc{\onehalf}{\frac{1}{2}}
\newc{\partialbar}{\bar{\partial}}
\newc{\psit}{\widetilde{\psi}}
\newc{\Tr}{\mbox{Tr}}
\newc{\ev}{\;\mathrm{eV}}
\newc{\mev}{\;\mathrm{MeV}}
\newc{\gev}{\;\mathrm{GeV}}

\def\chii0{\chi_i^0}
\def\chij0{\chi_j^0}

\newcommand{\bi}{\begin{itemize}}
\newcommand{\ei}{\end{itemize}}

\newcommand{\gsim}{\lower.7ex\hbox{$\;\stackrel{\textstyle>}{\sim}\;$}}
\newcommand{\lsim}{\lower.7ex\hbox{$\;\stackrel{\textstyle<}{\sim}\;$}}

\def\GeV{~{\mbox{GeV}}} 

\newcommand{ \slashchar }[1]{\setbox0=\hbox{$#1$}   
   \dimen0=\wd0                                     
   \setbox1=\hbox{/} \dimen1=\wd1                   
   \ifdim\dimen0>\dimen1                            
      \rlap{\hbox to \dimen0{\hfil/\hfil}}          
      #1                                            
   \else                                            
      \rlap{\hbox to \dimen1{\hfil$#1$\hfil}}       
      /                                             
   \fi}   
\def\misspt{{\slashchar p_T}}

\begin{document}

\preprint{MCTP-11-04}
\title{Searching for Top Flavor Violating Resonances}
\author{
Moira I. Gresham, Ian-Woo Kim and Kathryn M. Zurek \\
{\it Michigan Center for Theoretical Physics, \\ Department of Physics, University of Michigan, Ann Arbor, MI 48109 }
}
\date{\today}

\begin{abstract}
We study new top flavor violating resonances that are singly produced in association with a top at the LHC. Such top flavor violating states could be responsible for the Tevatron top forward-backward asymmetry.  Since top flavor violating states can directly decay to a top (or anti-top) and jet, and are produced in conjunction with another (oppositely charged) top, the direct signature of such states is a $t j$ (or $\bar{t} j$) resonance in  $t\bar{t}j$ events.  In general, these states can be very light and have ${\cal O}(1)$ couplings to the top sector so that they are copiously produced.  
We present a search strategy and estimate the discovery potential at the early LHC by implementing the strategy on simulated data.
For example, with 1 fb$^{-1}$ at 7 TeV,  we estimate that a $W'$ coupling to $d_R\bar{t}_R$ can be constrained at the $3 \sigma$ level for $g_R = 1 $ and $m_{W'} = 200 \mbox{ GeV}$, weakening to $g_R = 1.75$ for $m_{W'} = 600$ GeV.  With the search we advocate here, a bound at a similar level could be obtained for top flavor violating $Z'$s, as well as triplet and sextet diquarks. 

\end{abstract}

\maketitle


\section{Introduction}

Compared to the lighter quarks, the top remains relatively unconstrained.  In many models of new physics, such as Warped Extra Dimensions, Little Higgs, and Technicolor, the top sector is treated differently than the light quark sector on account of its strong coupling to electroweak symmetry breaking.  Thus many searches for new physics focus on 
discovering new physics in the top sector.  However, with ${\cal O}(1000)$ $t\bar{t}$ pairs at the Tevatron, in many cases  not enough statistics has been accumulated 
to be able to set strong limits on the couplings of the top to other SM particles.
Precise measurements of the top pair production cross-section \cite{d0tinvariantmass,cdftinvariantmass} and the few single top events  \cite{singletop} at the Tevatron serve to provide some constraints, though these are comparatively weak.  The constraints on $t_R$ couplings are especially weak, and the top CKM matrix elements are relatively unconstrained.

Recently, there has been an apparent anomaly in the top sector: the observation by the CDF experiment of a top forward-backward asymmetry ($A_{FB}^t$)~\cite{Aaltonen:2011kc}.  Although not conclusive, $A_{FB}^t$ as measured by CDF deviates from the SM theory prediction \cite{Kuhn:1998jr,Kuhn:1998kw,Bowen:2005ap,Almeida:2008ug}, especially at high invariant mass, $M_{t\bar{t}} > 450 \mbox{ GeV}$,  by more than $3\sigma$.  D0 also observes a larger than predicted asymmetry, but with much lower significance \cite{Abazov:2007qb, D0afb}.  

From the theoretical point of view, the difficulty with reproducing the observation is in generating the large asymmetry without disturbing the total cross-section or observed invariant mass spectrum of $t\bar{t}$ production, which is in agreement with the SM. 
The models generating the observed Tevatron $A^{t}_{FB}$ at tree level mainly fall into two categories: (i) $s$-channel exchange of vector mediators, the couplings of which to the top and the light quarks are axial and of opposite sign \cite{Sehgal:1987wi,Bagger:1987fz,Ferrario:2009bz,Frampton:2009rk,Chivukula:2010fk,Djouadi:2009nb,Bauer:2010iq,Alvarez:2010js},\footnote{There could be also a sizable $s$-channel contribution through vector operator interactions in unparticle theories~\cite{Chen:2010hm}.} or (ii) $t$-channel exchange of flavor-violating mediators such as new $W'/Z'$ gauge bosons \cite{Cheung:2009ch,Barger:2010mw,Cheung:2011qa,Jung:2009jz,Xiao:2010hm} or colored particles \cite{Shu:2009xf,Dorsner:2009mq,Arhrib:2009hu}. For various models of types (i) and (ii), comparative studies can be found in the literature~\cite{Cao:2009uz,Cao:2010zb,Choudhury:2010cd,Jung:2009pi,Jung:2010ri,Bai:2011ed}. Generally, $t$-channel exchange models can have an advantage over $s$-channel exchange models in explaining a large measured value of $A^t_{FB}$ in that $t$-channel models have mild additional contributions to the $t \bar{t}$ cross section, unlike resonant contributions from the $s$ channel.  

\begin{figure}
\centering
\subfigure[~$t\bar{t}$ production]{\includegraphics[width=0.2\textwidth]{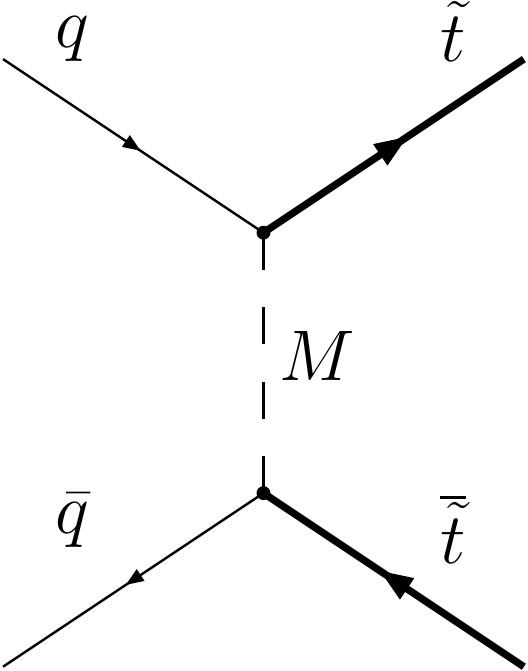}}\qquad
\subfigure[~t-channel]{\includegraphics[width=0.2\textwidth]{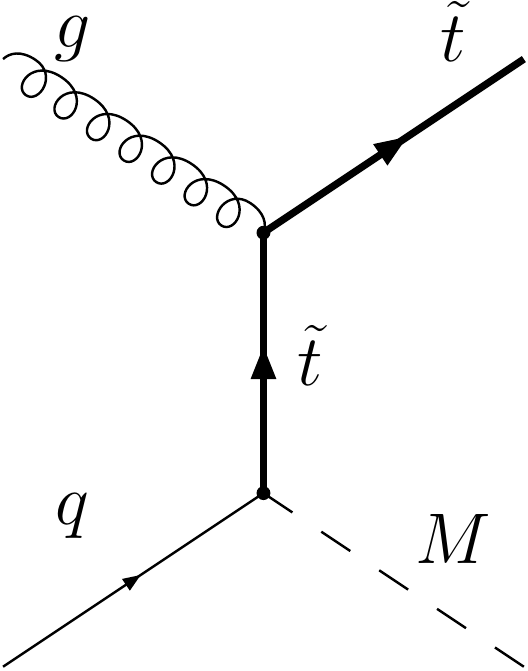}}~~~
\subfigure[~s-channel]{\includegraphics[width=0.2\textwidth]{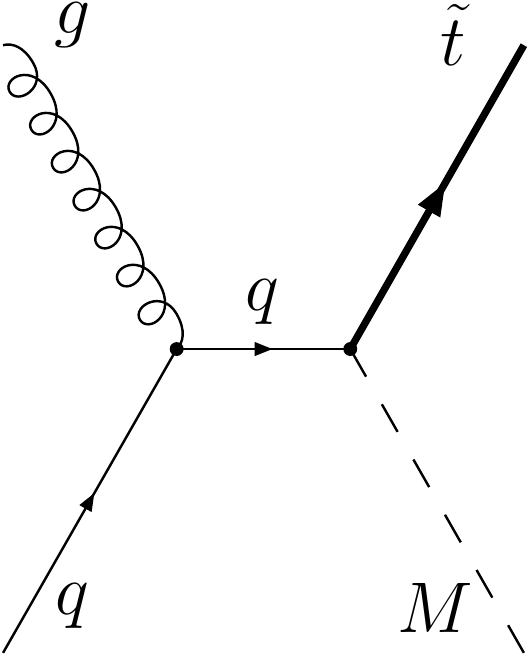}}~~~
\subfigure[~u-channel \qquad (M=$\phi^a$ only)]{\includegraphics[width=0.2\textwidth]{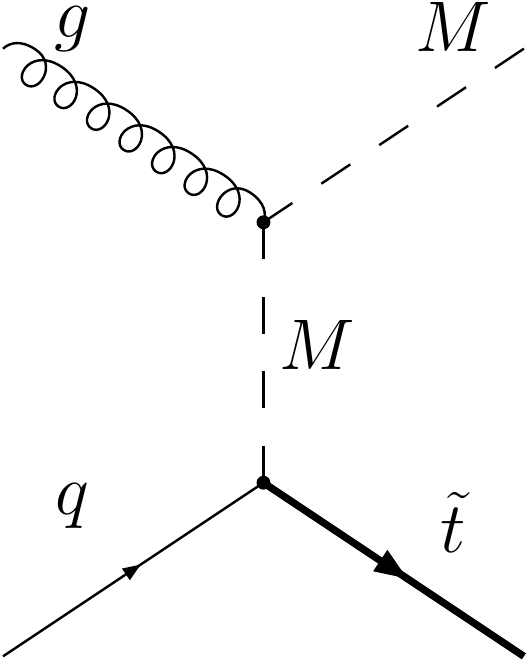}}
\caption{Tree level $t \bar{t}$ and single M production diagrams involving the mediator $M$ and the coupling $g_M$. The top quark, $\tilde{t}$, is $\tilde{t} = t$ when $M = W', Z'_H$ , and $\tilde{t} = \bar{t}$ when $M = \phi^a$  (triplet or sextet).}
\label{phiproduction}
\end{figure}

Since such $t$-channel exchange models have top flavor violation in mediator interactions, we expect a unique resonance signature. The logic is as follows:
\begin{itemize}
\item 
Top forward-backward asymmetry generating models of type (ii) discussed above have interactions of the form $g M \bar{t} q$ where $M$ is the mediator, $q$ is a light quark, and $g$ is order 1. Thus the production of $M$ through $q g \rightarrow M t$ as in Fig.~\ref{phiproduction} is expected to be substantial.
\item 
For mediators with mass $m_M > m_t$, this implies $M$ can decay through $M \rightarrow \tilde{t} q$, where $\tilde{t} = t~\text{or}~\bar{t}$.
Therefore, a $\tilde{t} j$ resonance should exists in $\tilde{t} \tilde{t} j$ events, where $j$ is a jet formed from the light quark $q$. 
\item 
To avoid constraints from same sign top pair production, we assume that $M$ is not self-conjugate, and then the signature is a top-jet ($t j$) or anti-top-jet ($\bar{t} j$) resonance in $t \bar{t}$ plus jet events. 
\item 
Due to baryon number conservation, the final state light quark baryon number must match that of the initial state quark.
In a $p p $ machine (as opposed to $p \bar{p}$), which has quark collisions dominantly over anti-quark collisions,  
the resonance will be dominantly either $t j$ or $\bar{t} j$, depending on the baryon number of the mediator, $B_M = \pm 2/3$  or $B_M = 0$, respectively.

\end{itemize}

Therefore, in contrast to other LHC search studies for models related to the $A_{FB}^t$ anomaly, which have focused on the $t \bar{t}$ or dijet invariant mass distributions~\cite {Choudhury:2010cd,Morrissey:2009tf,Bai:2011ed},\footnote{For generic colored resonance search through QCD interations, see \cite{Han:2010rf}.} here we emphasize top-jet resonances at the LHC as the most direct evidence of top flavor violating physics. 
The relevance of 
top-jet resonances to a search for various top flavor violating models has been mentioned in previous studies \cite{Shu:2009xf,Dorsner:2009mq,Arhrib:2009hu}. In this paper, however, we consider top-jet resonances at the early LHC in depth.  We propose a systematic search strategy for such top-jet resonances. Using this strategy along with simulated data, we present the early LHC reach for models with top flavor violation. 

The early LHC analysis that we propose here essentially depends on only two considerations: the mass of the resonance and the partial production cross-section of the resonance in decay to top and jet.  A key part of the analysis is to look for $\tilde{t}j$ resonances in the {\em low to intermediate} invariant mass region ($\sim 200 - 600$ GeV).  Thus the method employed is quite different from the usual bump hunting strategies, which fit the background at low invariant mass and then search for a resonance at high invariant mass.  Here, the continuum background should be fit at intermediate to high invariant mass, and an excess of events searched for at low to intermediate invariant masses. Considering an additional angular variable---the angle between top and jet in the lab frame---also helps in separating signal from background in the low invariant mass region.  We will show in this paper that this type of low to intermediate mass resonance can be successfully extracted from background $t\bar{t} j$ processes.

The outline of this paper is as follows.  In Section \ref {sec:2}, we will introduce top-flavor violating models and benchmark points for analysis. In Section \ref{sec:3}, we review the reconstruction scheme for tops and build a search strategy for top-jet resonances. We then show results of event generation and detector simulation. Finally, in Section \ref{sec:4} we present the reach potential of this search in the early phase of the LHC operation and conclude in Section \ref{sec:5}.

\section{Models of Top Flavor Violation}\label{sec:2}

We classify models with top flavor violation according to the following properties of the new top-flavor changing particle $M$: 
\begin{itemize}
\item Spin : vector or scalar; 
\item Color representation : singlet, octet, triplet or sextet;  
\item Right-handed isospin : changing or neutral,  
\end{itemize} 
where right-handed isospin means isospin assigned to right-handed quarks in the same way as left-handed quark doublets. 
To avoid severe experimental constraints, we require that baryon number not be violated by $M$. For the purposes of this analysis, the most dominant coupling of the new particle, $M$, with the standard model involves only the top quark and light quarks $u$ and $d$. Hence, the relevant interaction of $M$ with the standard model is fully determined by specifying the above characteristics. 

In this paper, we will quote results for four classes of models: $W'$, $Z'_H$, color triplets and sextets.  The $W'$ will take a down quark to a top quark, the $Z'_H$ takes up to top, and the others take an up quark to an anti-top quark.  The important interaction Lagrangians for these models are:
\begin{eqnarray}
{\cal L}_{\rm W'} &=& {1 \over \sqrt{2}} \bar{d} \gamma^\mu g_R P_R t W'_\mu + \mbox{h.c.}  \nonumber \\ 
{\cal L}_{\rm Z'_H} &=& {1 \over \sqrt{2}} \bar{u} \gamma^\mu g_R P_R t {Z'_H}_\mu + \mbox{h.c.} \nonumber \\ 
{\cal L}_\phi &=&   \bar{t^c} T^a_r  (g_L P_L + g_R P_R) u \phi^a+ \mbox{h.c.},
\label{lagrangian}
\end{eqnarray}
where $\phi^a$ is a color triplet or sextet, and $T_r^a$ is the generator of the SU(3) representation.
We consider only $Z'$s that are gauge bosons of a horizontal symmetry---say, an $SU(2)_H$ connecting the first and third generations---which carry a horizontal charge and are thus not self-conjugate.   A similar but not identical analysis could be carried out for self-conjgate $Z'$s, which can go both to $tj$ and $\bar{t}j$ resonances.

Fig.~\ref{phiproduction}a shows the $t$-channel Feynman diagram of $t\bar{t}$ production that contributes to the total cross section and $A_{FB}^t$ at the Tevatron.  Typically, in these models, the best fit values of the $M$ mass, $m_M$, are within 200~{\rm GeV} -- 1000~\GeV, with the required coupling to produce the observed asymmetry ranging from $g_M \sim 1$ for a 200 GeV mediator to $g_M \sim 10$ for a 1 TeV mediator.  For the purposes of this analysis, we restrict ourselves to mediators having perturbative couplings, and masses consistent with those considered in the literature, ranging from a 200 GeV $W'$ or $Z'$ to a 400 or 600 GeV color triplet or sextet.

At the early LHC energies of $\sqrt{s} = 7$ TeV, $M$ can be produced singly as in Figs.~\ref{phiproduction}b-d, or produced doubly through SM gauge interactions and top flavor violating interactions.  For mediators at least moderately heavier than the top quark, the dominant production will be a single resonance along with a top, with di-resonance production significantly suppressed for a center of mass energy $\sqrt{s} = 7 \mbox{ TeV}$.\footnote{ Di-resonance production is only very important for the 200 GeV scalar sextet model, as shown in Table~\ref{cross-sections}.}  The tree-level single and double production cross-sections for several top-flavor-violating models are listed in Table~\ref{cross-sections}. Since these models have large couplings, substantial higher-order corrections to the cross-sections should be expected.

\begin{table}
\begin{tabular}{|c|c|c|c|c|c|c|}
\hline
Mediator   &\multicolumn{4}{|c|}{Tree Level Single (Double)} & {Semileptonic} & Reconstruction  \\ 
Mass & \multicolumn{4}{c|}{Production Cross-Section (pb)} & {cut efficiency} & efficiency    \\ \cline{2-5} 
& $W'$ & $Z'$ & $\bar{\bf 3}$ & ${\bf 6}$ &   & \\ \hline \hline
200 \mbox{ GeV} & 40 (4.7) &  75 (8.3) & 44 (7.6)& 149 (134) & 1.3 \%  & 36 -- 39 \%  \\ \hline
300 \mbox{ GeV} & 14 (0.8) & 27 (1.5) & 17 (0.9)& 54 (15)& 3.2 \% & 34 -- 36 \% \\ \hline
400 \mbox{ GeV} & 5.7 (0.2) & 12 (0.4) & 7.7 (0.2) & 23 (2.6)& 3.5 -- 3.8 \%	& 24 -- 25 \% \\ \hline
600 \mbox{ GeV} & 1.3 (0.02) & 2.9 (0.04) & 2.0 (0.01)& 5.6 (0.17)& 3.9 -- 4.2 \% & 18 -- 19 \% \\ \hline
\end{tabular}
\caption{The total tree level mediator plus top production cross-section and efficiency after cuts. 
The cross-sections are calculated taking $g_R = 1$ in the Lagrangian of Eq.~(1).  The semi-leptonic cuts correspond to requiring a single muon or electron, missing energy, a tagged $b$-jet and 3 additional jets with rapidities and $p_T$ described in Sec.~(\ref{Sec:signatures}).  The reconstruction efficiency for tops after applying these cuts (using the method outlined in appendix A) is also listed. The efficiencies  depend essentially only upon the resonance mass. The efficiency numbers quoted were obtained from analyzing $W'$, $Z'_H$, and triplet models. }
\label{cross-sections}
\end{table}

\section{Model Independent Analysis of Top-Jet Resonance}\label{sec:3}

We now set up the search method for top-jet resonances at the LHC. 
After the decay of $M$, the final states in the signal are decay products of a $t\bar{t}$ pair and one quark jet. In the following sections, we propose a procedure for reconstructing tops and identifying signatures. We also present simulation results for the benchmark cases included in Table \ref{cross-sections}.
 
\subsection{Top-Jet Resonance Signature}
\label{Sec:signatures}

The signatures depend primarily on the $t\bar{t}$ decay modes:\footnote{In this paper, ``lepton'' is taken to be electron or muon.} (a) hadronic mode ($t\bar{t}\rightarrow b\bar{b} j j j j $), (b) semi-leptonic mode ($t\bar{t} \rightarrow b\bar{b} j j \ell \nu $), (c) leptonic mode ($t\bar{t} \rightarrow b \bar{b} \ell^+ \ell^- \nu \bar{\nu} $) and (d) decay modes involving taus.  In this analysis, we consider only the semi-leptonic decay mode.   The reason for this is as follows.  In fully leptonic events, we cannot fully reconstruct the top-quark momentum event by event.  And in fully hadronic modes there is a high multiplicity of jets, leading to large uncertainties, misidentification, and combinatoric issues. Furthermore, we use the sign of the lepton to distinguish between $t$ and $\bar{t}$ in each event (assuming negligible like-sign top production). The branching fraction for $t\bar{t}$ to decay semi-leptonically is $\sim30\%$. 

Including a jet from $M \to \tilde{t}j$, we have two $b$ jets, three non-$b$ jets, one lepton, and missing energy in the signal. 

To select the $t\bar{t} j$ signal, we require: 
\begin{itemize}
\item Exactly one electron or muon with $p_T > 20$ GeV and $|\eta| < 2.5$.
\item Photon and $\tau$ veto.
\item At least five jets with $p_T > 20$ GeV and $|\eta| < 3.0$, with at least one of the jets having a $b$-tag.  Four of the jets are decay products of the tops, with the additional jet coming from the mediator decay to $\tilde{t}j$.
\item $E_T^{miss} > 20$ GeV. 
\end{itemize}
These cuts are chosen in accordance with those for $t\bar{t}$ analyses as laid out in the CMS  Technical Design Report~\cite{Ball:2007zza}.  

To find the $\tilde{t} j$ resonance and to further reduce background, it is important to identify top and anti-top pairs out of the multi-jet, lepton, and missing $E_T$ signature. We do a $\chi^2$-based cut on the lepton and jet kinematics to the $t \bar{t}$ hypothesis,
\begin{itemize}
\item $\chi^2_{t \bar{t}} / \text{d.o.f.} \leq 1$,
\end{itemize}
as detailed in Appendix~\ref{top reconstruction}.  The main background after these cuts is SM $t\bar{t}$ pair production with additional jets.  

\begin{figure}
{\centering
\subfigure[~Invariant Mass]{\includegraphics[width=0.49\textwidth]{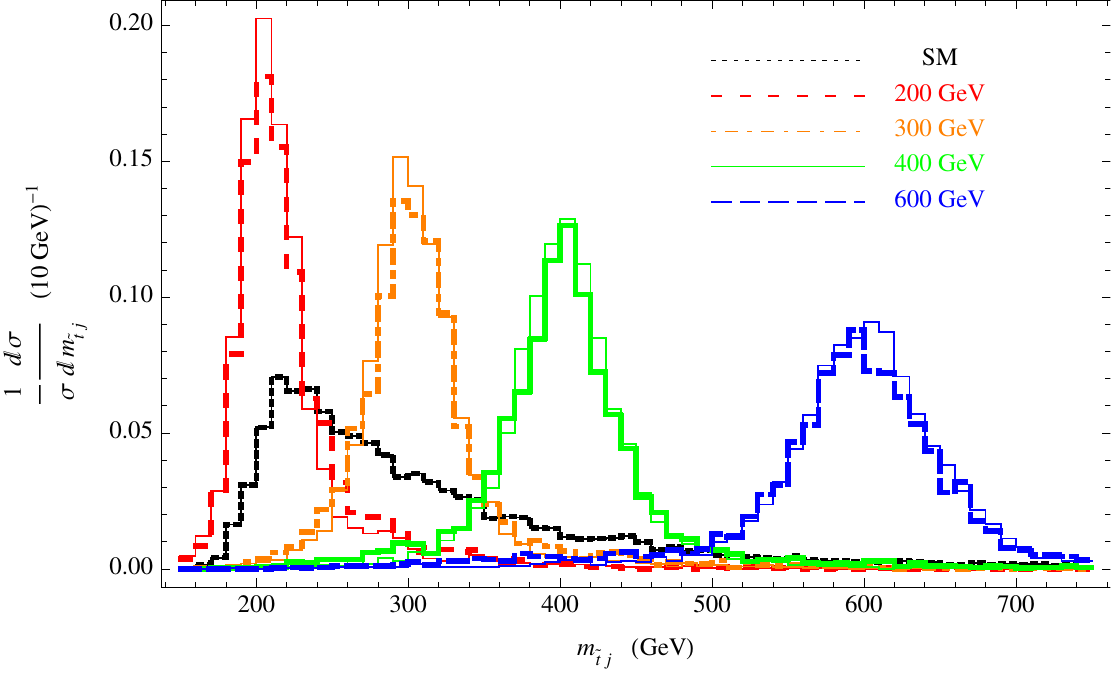}}~
\subfigure[~$\cos \theta_{\tilde{t}j}$]{\includegraphics[width=0.49\textwidth]{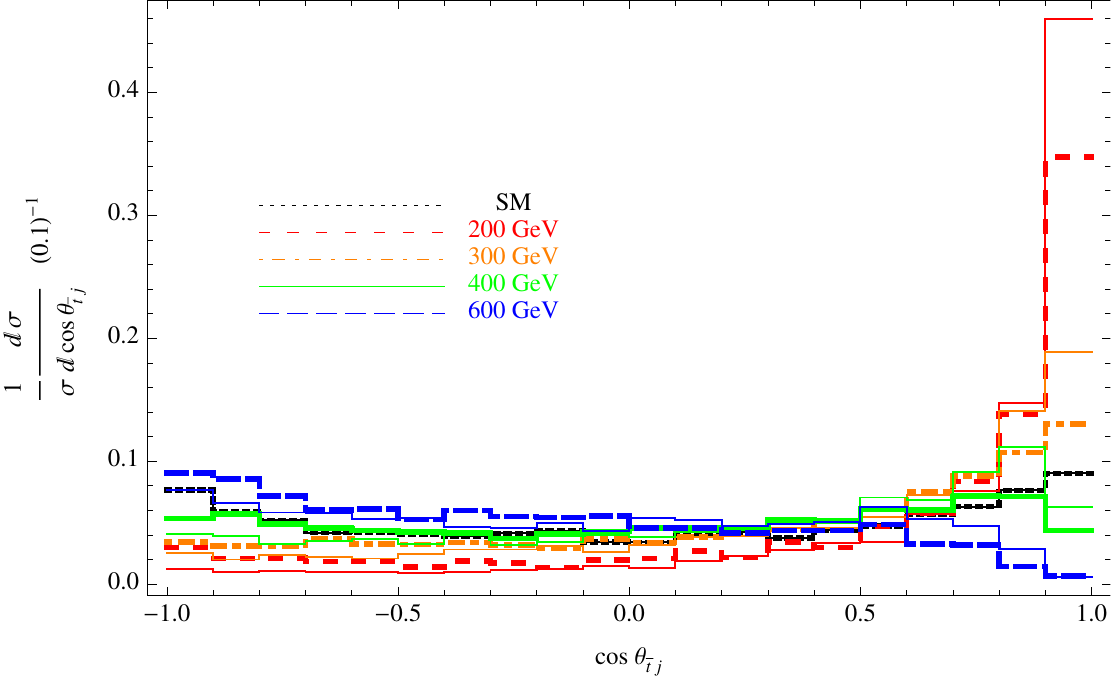}\vspace{.25cm}}
\caption{Left: ${ 1 \over \sigma}{d\sigma \over d M_{\tilde{t} j} }$ distribution of reconstructed tops and one extra jet in units of $(10~\text{GeV})^{-1}$ for smeared parton-level signal events and for Standard Model (SM) simulated reconstructed detector-level events. The thick lines are the distribution for a $W'$ resonance ($\bar{t} j$) and the thinner solid lines for a $Z'_H$ ($\bar{t} j$). Right: ${ 1 \over \sigma}{d\sigma \over d\cos \theta_{\tilde{t} j} }$ distribution of reconstructed tops and one extra jet  for the same events. 
 }\label{invariantmass}
 }
\end{figure}

 The invariant mass is the obvious variable to consider in a resonance search. In Fig.~\ref{invariantmass}a, we show the $m_{\tilde{t}j}$ distribution of several top-flavor-violating models along with the SM distribution. Here, to get a semi-realistic distribution, we smear visible particle momenta of parton-level events  according to the  LHC-motivated uncertainties given in Appendix \ref{CMS}, and then we reconstruct top particles.  
 The peaks in the $m_{\tilde{t}j}$ distribution for the signals are clear. However, the SM
 distribution can obscure low mass signals since it peaks around 200~GeV. 
 The precise location of the SM peak just above the top mass depends on the choice of $p_T$ cuts; the extra jet in Standard Model $t \bar{t} j$ events tends to be fairly soft, but still hard enough to have passed the cuts.
 
 Due to the overlap between the low mass resonance and the SM peaks, we consider another variable, $\cos \theta_{\tilde{t} j}$, defined as the cosine of the angle between the reconstructed (anti) top and the remaining jet in the lab frame.
This variable is closely connected with the velocity distribution of $M$, and thus has relatively small model dependencies since the Lorentz boost of $M$ in the lab frame is primarily determined by $m_M$. 
 As can be seen in Fig. \ref{invariantmass}b, 
 the angular variable can be an efficient separator for low mass resonances. For a low mass resonance near the top quark mass, the resonant particle will be fairly boosted in the lab frame at LHC energies, so that the resultant $\tilde{t}$ and $j$ are collimated. On the other hand, for high mass resonances we expect $\cos \theta_{\tilde{t} j} \approx -1$ since the resonant particle will be nearly at rest in the lab frame, resulting in back-to-back $\tilde{t} j$ decay. Compared to the signal, 
 the SM $\cos \theta_{\tilde{t}j}$ distribution is  relatively flat. 

We find that the variables $M_{\tilde{t} j}$ and $\cos \theta_{\tilde{t} j}$ considered together can efficiently separate signal from background. 
In the following analysis, we use the combined $\frac{d^2 \sigma}{d M_{\tilde{t}j} d \cos \theta_{\tilde{t}j}}$ for finding the highest significance over background in a given data set.  

In this paper we focus on single production of top-flavor-changing mediators. This focus is certainly justified for the 200+ GeV $Z'_H$, $W'$, and triplet models. (See Table~\ref{cross-sections}.) We leave a detailed study of the effect of double mediator production, especially in the sextet model (which has a large contribution from double production due to a large group theoretical factor) 
for future work. 
With double production contributions to the tree-level $200$~GeV $Z'_H$, $W'$, and triplet production cross-sections occurring at only the 10--20\% level, we expect that doubly-produced mediators would not change the analysis much even in these cases, and thus we present only the single production results here.

\subsection{Event Generation}

Standard model $t \bar{t}$ and $t \bar{t} j$ events corresponding to about $10~\text{fb}^{-1}$ at the $\sqrt{s} = 7$ TeV LHC  were generated as background. In order to avoid over- or under-counting of events, we generated the SM background events using {\tt MadGraph/MadEvent 4.4.32} \cite{Alwall:2007st} with matrix element / parton shower (ME/PS) matching in the MLM scheme~\cite{Caravaglios:1998yr}, which was implemented in the {\tt MadGraph/MadEvent} package using {\tt PYTHIA}~\cite{Sjostrand:2006za}. The events were generated using a fixed renormalization scale and factorization scale at 200~{\rm GeV}. The total cross section of the SM $t\bar{t}$ plus zero or one jet is given by $\sim$107 pb with {\tt QCUT = 30 } and {\tt xqcut = 20} in {\tt MadGraph} and {\tt PYTHIA} settings. 

For signals, we use a combination of {\tt MadGraph5 v0.6.1 / MadEvent 4.4.44}, which has an implementation of color exotic particles. We generate ten thousand single mediator production events,
\[
p p \rightarrow M \tilde{t} \rightarrow {t} \bar{{t}} j, \qquad p p \rightarrow \bar{M} \bar{\tilde{t}} \rightarrow  {t} \bar{{t}}  j ,
\]
with top pairs decayed semi-leptonically
 for each of the benchmark models: $W'$, $Z'_H$, triplet, and sextets with $g_R = 1$ (Eq.~\eqref{lagrangian}) and $m_M = 200,300,400, 600$ GeV, as listed in Table~\ref{cross-sections}.

To perform a semi-realistic analysis, we employ the {\tt PGS} detector simulator. Note that we do not apply a $K$-factor to the cross section we present in this paper.

\subsection{Top-Jet Resonance Search} 

One difference here, as compared to a typical new physics invariant mass bump hunt, is that we are looking for relatively {\em low mass} resonances.  As just discussed, low mass resonances will show up as a bump on top of the peaked region of the Standard Model distribution. Thus rather than fitting the background to an exponential or power law at low invariant mass and looking for excesses at high invariant mass, a different approach should be taken for the low mass ($\mathcal{O}(200~\text{or}~300~\text{GeV})$) mediators.  In these cases, we fit the background to the high invariant mass part of the distribution and look for excesses in the low invariant mass bins. Without knowing the mass of the resonance \emph{a priori}, one should start by normalizing background over a large mass range. In this case, the low-mass resonance will show up as a very small bump, with anti-bumps (deficits of observed events with respect to expected events) on either side. Such an anti-bump/bump/anti-bump pattern is an indication that background should be fit in a region away from the bump.   

In Figs.~{\ref{200GeVInvMass}--\ref{600GeVInvMass},
 we show $m_{\tilde{t} j}$ and $\cos \theta_{\tilde{t} j}$ distributions from simulated events within a $W'$ model, with appropriately normalized expected background shown as a solid blue histogram. The errors are taken to be purely statistical and are not shown.  The background normalization is fixed so that the total ``measured'' events in a certain mass range---say, 400 to 700 GeV---matches the number of expected background events in that same range. Invariant mass bumps show up clearly, and bumps in the $\cos \theta_{\tilde{t} j}$ are also apparent for the lower mass resonances. Bin sizes were set at $\Delta m_{\bar{t} j} = 50$ GeV and $\Delta \cos \theta_{\bar{t} j} = 0.5$ in all of Figs.~~{\ref{200GeVInvMass}--\ref{600GeVInvMass}.

In order to take advantage of both the $m_{\tilde{t} j}$ and $\cos \theta_{\tilde{t} j}$ distributions, we consider the two-dimensional $\frac{d^2 \sigma}{d M_{\tilde{t}j} d \cos \theta_{\tilde{t}j}}$ distribution. Expected background is normalized by using the procedure above. Bin size is then varied over $M_{\tilde{t} j}$ and  $\cos \theta_ {\tilde{t} j} $ so as to maximize the $\chi^2$ of the two-dimensional distribution in a single bin, increasing the significance of the signals. At least for low-mass resonances, the significance is enhanced by considering the two-dimensional distribution. This is demonstrated in Figs.~\ref{contour plots} and \ref {best bins}. Fig.~\ref{contour plots} is a density plot of the two-dimensional differential distribution for simulated standard model and for signal only events, showing how the distributions shift not only in top-jet invariant mass, but also $\cos \theta_{\tilde{t}j}$, relative to the SM distribution.\footnote{Signal and background events can be considered separately due to the negligible interference between standard model and signal events.}  Through consideration of these two-dimensional distributions, the bin in $M_{\tilde{t} j}$ and  $ \cos \theta_{\tilde{t} j}$ that gives the largest signal significance can be found.  This is shown in Fig.~\ref {best bins} for a couple of the resonances considered. The $\chi^2$ values for the $W'$ model in Fig.~\ref{best bins} should be contrasted with the $\chi^2$ values shown in Figs.~{\ref{200GeVInvMass}--\ref{600GeVInvMass}. For example, the maximum $\chi^2$ given the $50$ GeV invariant mass binning is just above $5$ for the 300 GeV case (Fig.~\ref{300GeVInvMass}), whereas a two-dimensional optimal binning for the same data gives $\chi^2 = 7.3$ (Fig.~\ref{best bins}). 

Two-dimensional binning is less advantageous (and also limited by statistics) for the higher mass resonances. 
The optimization procedure we used in determining the appropriate bin is as follows: We varied the number of equal-sized bins from 1 to 24 over the invariant mass range from $150$ GeV to $750$ GeV and from 1 to 10 over the entire $\cos \theta_{\tilde{t} j}$ range. The $\chi^2$ for each bin was calculated for all of the bin size combinations. The optimal bin size is the one that gives the largest $\chi^2$ in a single bin. We set $\chi^2 = 0$ for a given bin if there were less than five expected background events in that bin, to avoid ambiguities in the interpretation of the significance associated with a given $\chi^2$. 

The slight differences between the differential distributions of the different models (Fig.~\ref{contour plots}) is due to the somewhat different production mechanisms of the models. First, the $W'$ couples to $d g$ in the initial state while $Z'_H$ and $\phi^a$ couple to $u g$, which lead to different center-of-mass energy distributions. Additionally, the colored particles can be singly produced in association with a top through an additional important $u$-channel diagram (Fig.~\ref{phiproduction}d), as compared to the color singlet $W'$ and $Z'_H$. This causes a slight difference in event momentum distributions, which in turn affects the reconstruction and angular distributions. 
 However,  despite the different production mechanisms, the signatures of the various models are more-or-less the same for a given mediator mass. We will see that the LHC reach for the various models is about the same for a given mediator mass.

\begin{figure}
\includegraphics[width=1\textwidth]{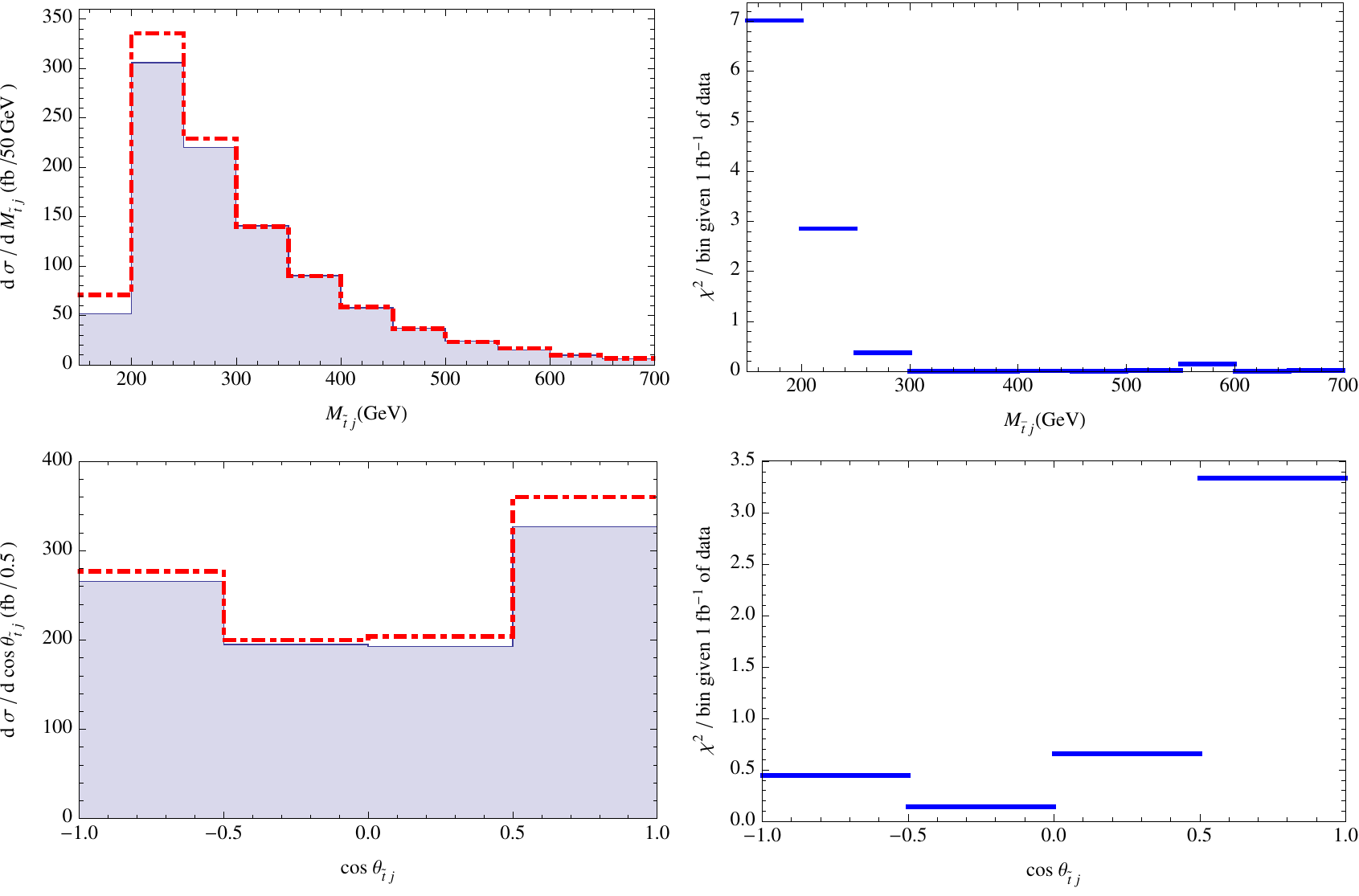}\\
{\bf 200 GeV Resonance}
\caption{Differential cross section in fb versus invariant mass (left top) or $\cos \theta_{\bar{t} j}$ (left bottom) and corresponding $\chi^2$ per bin (right) for the background only hypothesis given 1 $fb^{-1}$ of data (right) and a 200 GeV $W'$ resonance with coupling $g_R = 1$. The normalized background is shown in solid blue, and the ``measured'' (Standard Model + signal) differential cross-section is shown as a red dashed-dotted line. The overall background normalization was fixed by matching to the ``measured'' cross section between invariant masses from 300 to 700 GeV.}
\label{200GeVInvMass}
\end{figure}

\begin{figure}
\includegraphics[width=1\textwidth]{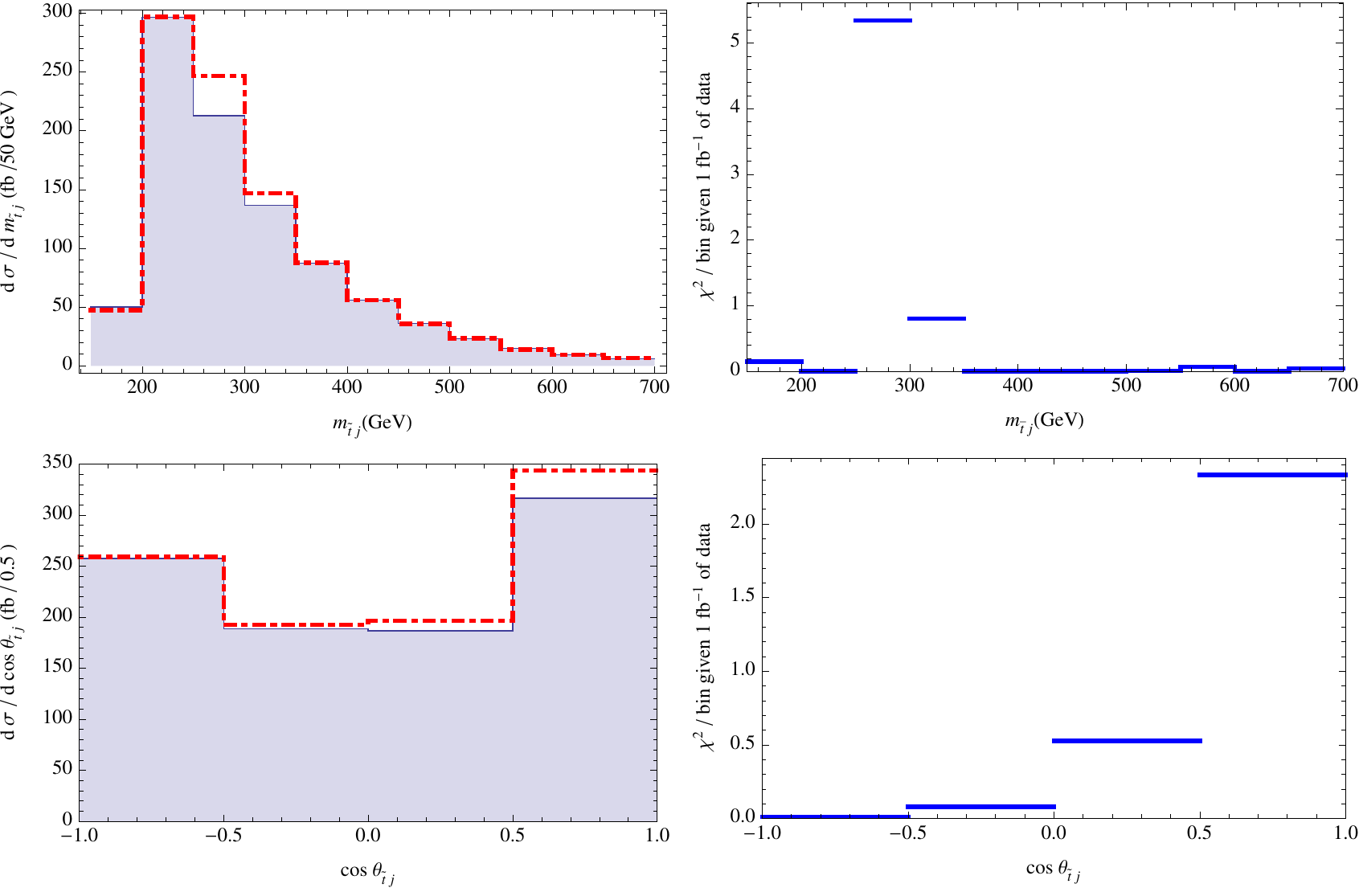}\\
\vspace{.25cm}
{\bf 300 GeV Resonance}
\caption{Differential cross section in fb versus invariant mass (left top) or $\cos \theta_{\bar{t} j}$ (left bottom) and corresponding $\chi^2$ per bin (right) for the background only hypothesis given $1~\text{fb}^{-1}$ of data (right) and a 300 GeV $W'$ resonance with coupling $g_R = 1$. The overall background normalization was fixed by matching to the ``measured'' cross section between invariant masses from 400 to 700 GeV.}
\label{300GeVInvMass}
\end{figure}

\begin{figure}
\includegraphics[width=1\textwidth]{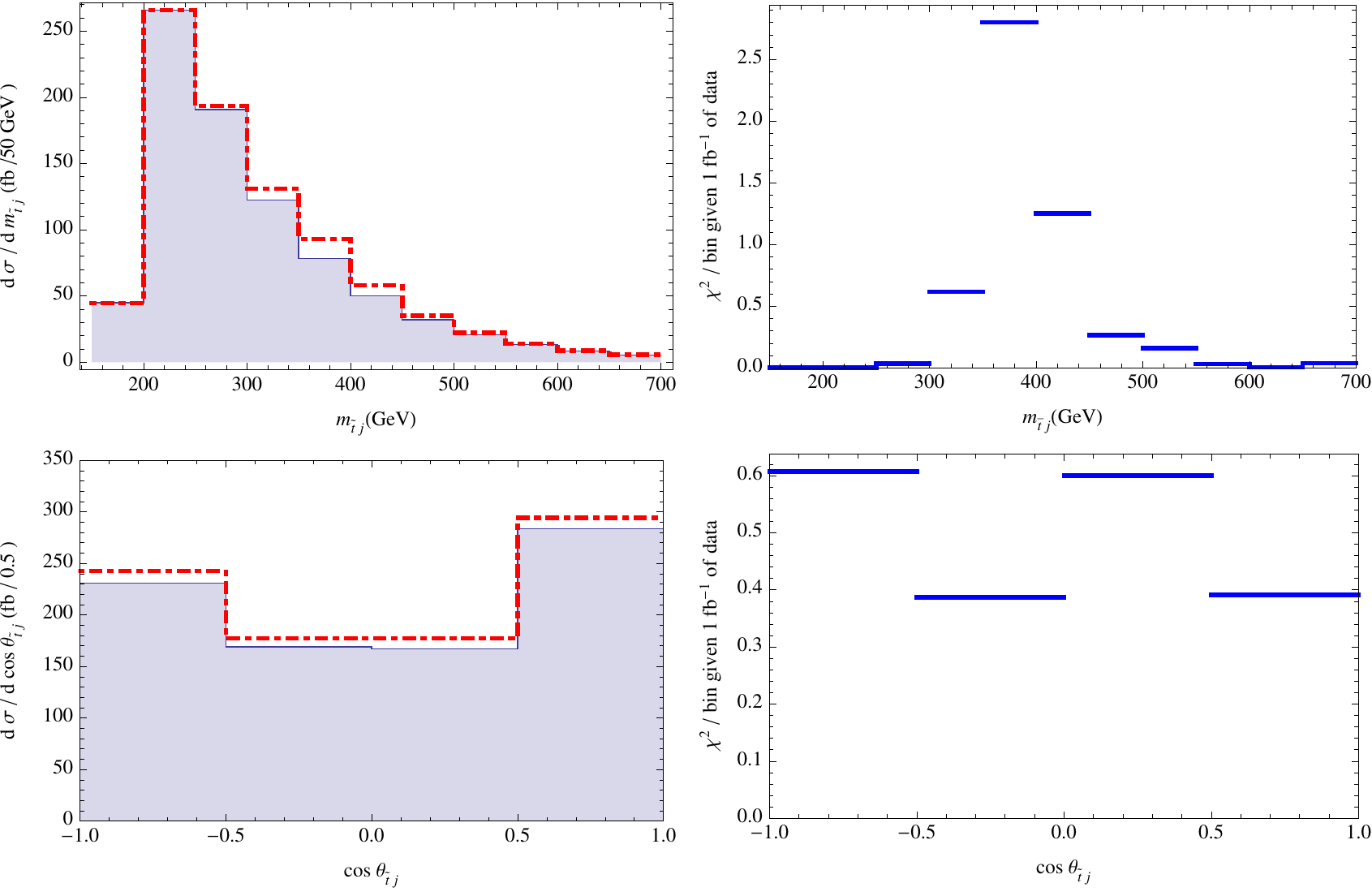}\\
\vspace{.25cm}
{\bf 400 GeV Resonance}
\caption{Differential cross section in fb versus invariant mass (left top) or $\cos \theta_{\bar{t} j}$ (left bottom) and corresponding $\chi^2$ per bin (right) for the background only hypothesis given $1~\text{fb}^{-1}$ of data (right) and a 400 GeV $W'$ resonance with coupling $g_R = 1$. The overall background normalization was fixed by matching to the ``measured'' cross section between invariant masses from 150 to 250 GeV.}
\label{400GeVInvMass}
\end{figure}

\begin{figure}
\includegraphics[width=1\textwidth]{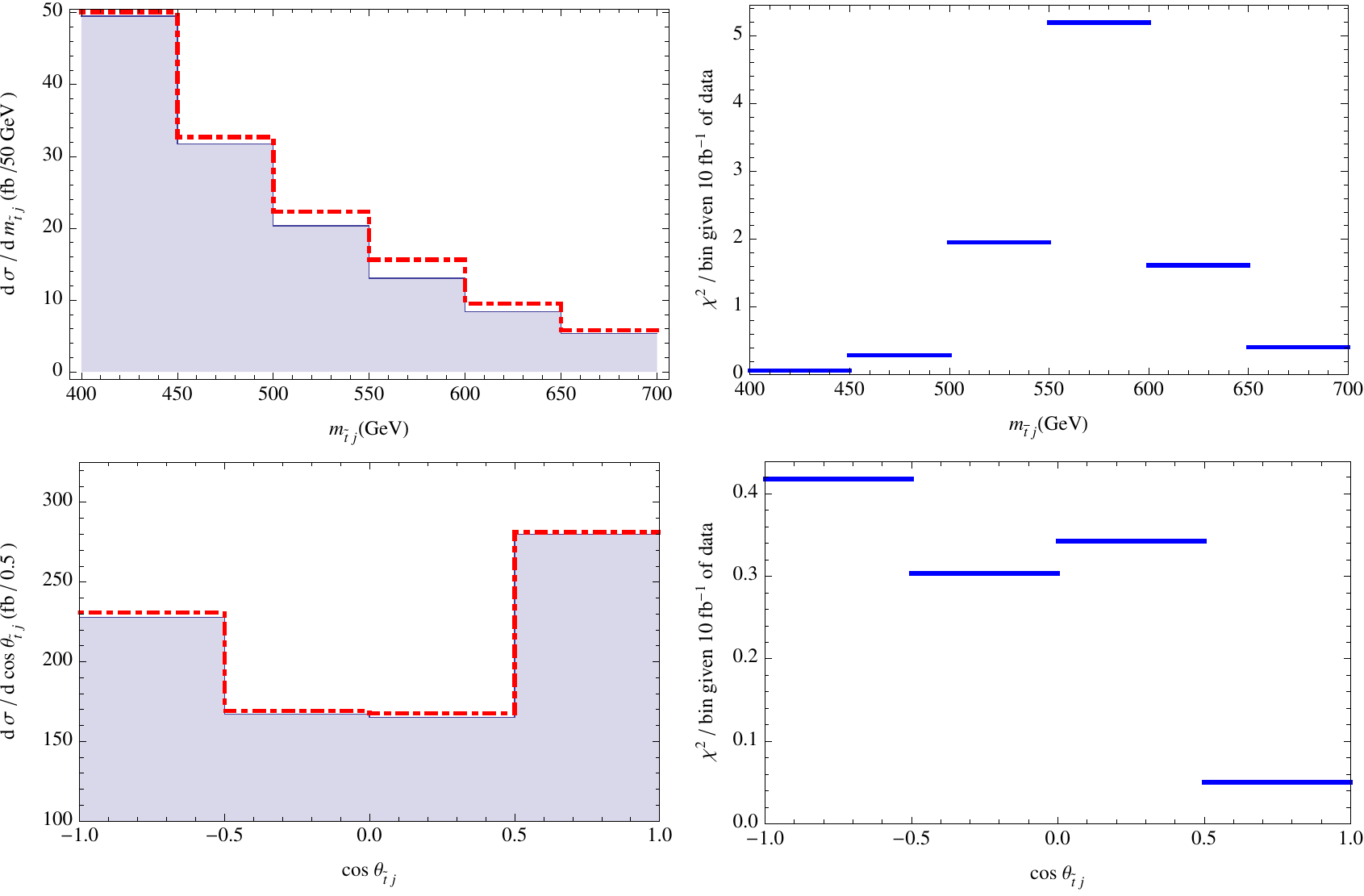}\\
\vspace{0.25 cm}
{\bf 600 GeV Resonance}
\caption{Differential cross section in fb versus invariant mass (left top) or $\cos \theta_{\bar{t} j}$ (left bottom) and corresponding $\chi^2$ per bin (right) for the background only hypothesis given 10 fb$^{-1}$ of data (right) and a 600 GeV $W'$ resonance with coupling $g_R = 1$. The overall background normalization was fixed by matching to the ``measured'' cross section between invariant masses from 150 to 300 GeV.}
\label{600GeVInvMass}
\end{figure}

\begin{figure}
\centering
{\bf $W'$:}\\ 
\includegraphics[width = 0.9 \textwidth]{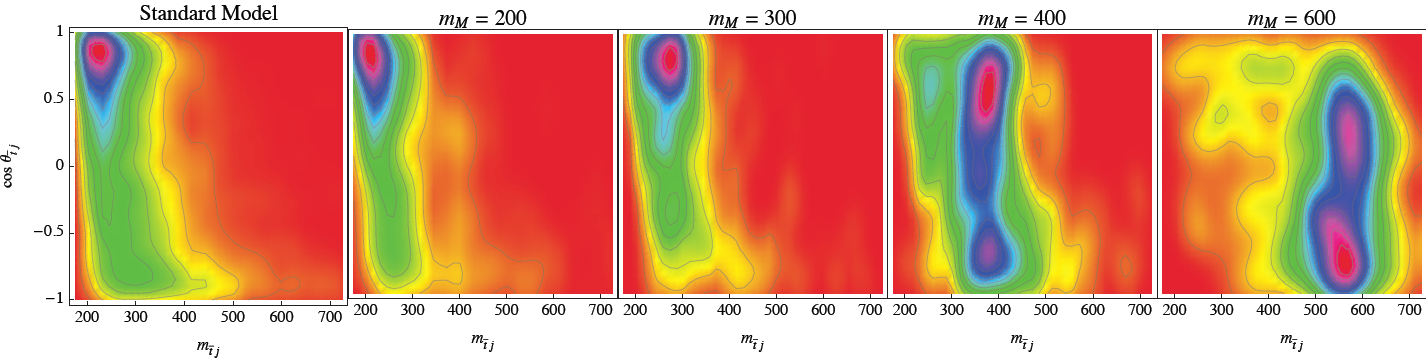}\\
{\bf $Z'_H$:}\\ 
\includegraphics[width = 0.72 \textwidth]{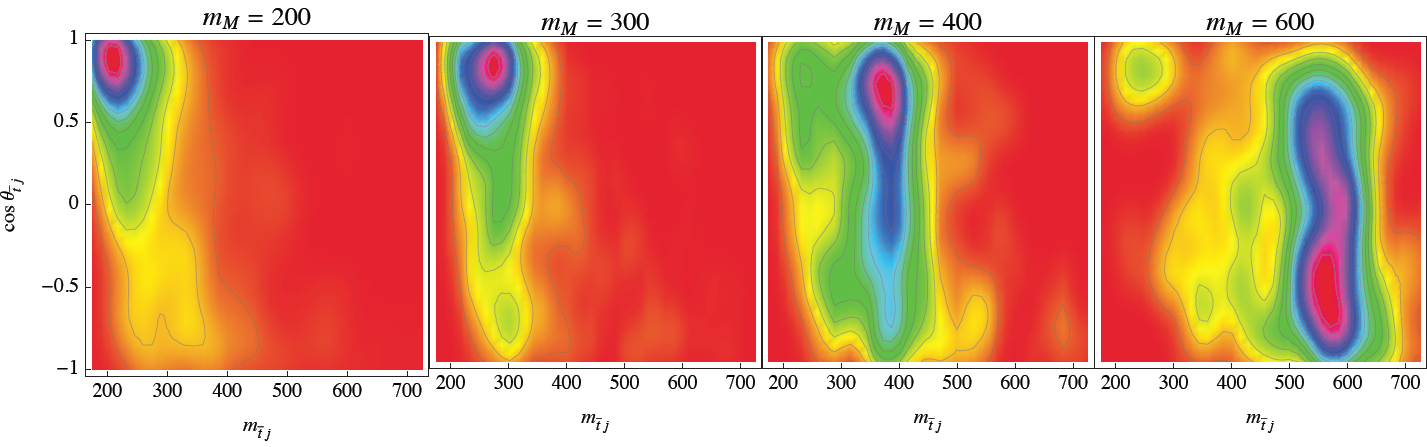}\\
{\bf triplet:}\\ 
\includegraphics[width = 0.9 \textwidth]{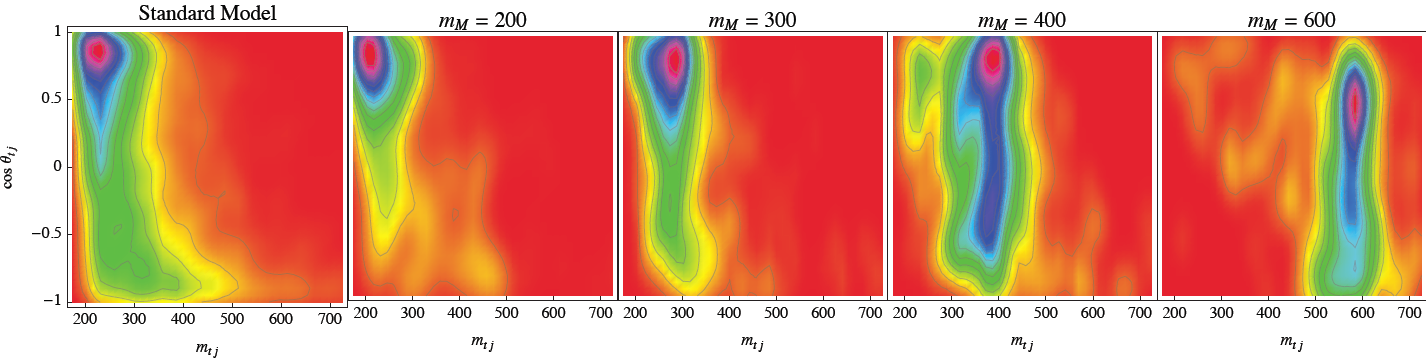}\\
\caption{${1 \over \sigma }{d^2 \sigma \over d m_{\bar{t} j} d \cos \theta_{\bar{t} j}}$ and ${1 \over \sigma }{d^2 \sigma \over d m_{t j} d \cos \theta_{t j}}$  distributions of reconstructed top-jet pairs from  simulated detector-level Standard Model (left top, left bottom) and signal only $W'$ ($\bar{t} j$ resonance), $Z'_H$ ($\bar{t} j$ resonance), and triplet ($t j$ resonance) events (right).  }
\label{contour plots}
\end{figure} 

\begin{figure}
\centering
\includegraphics[width = 0.4 \textwidth]{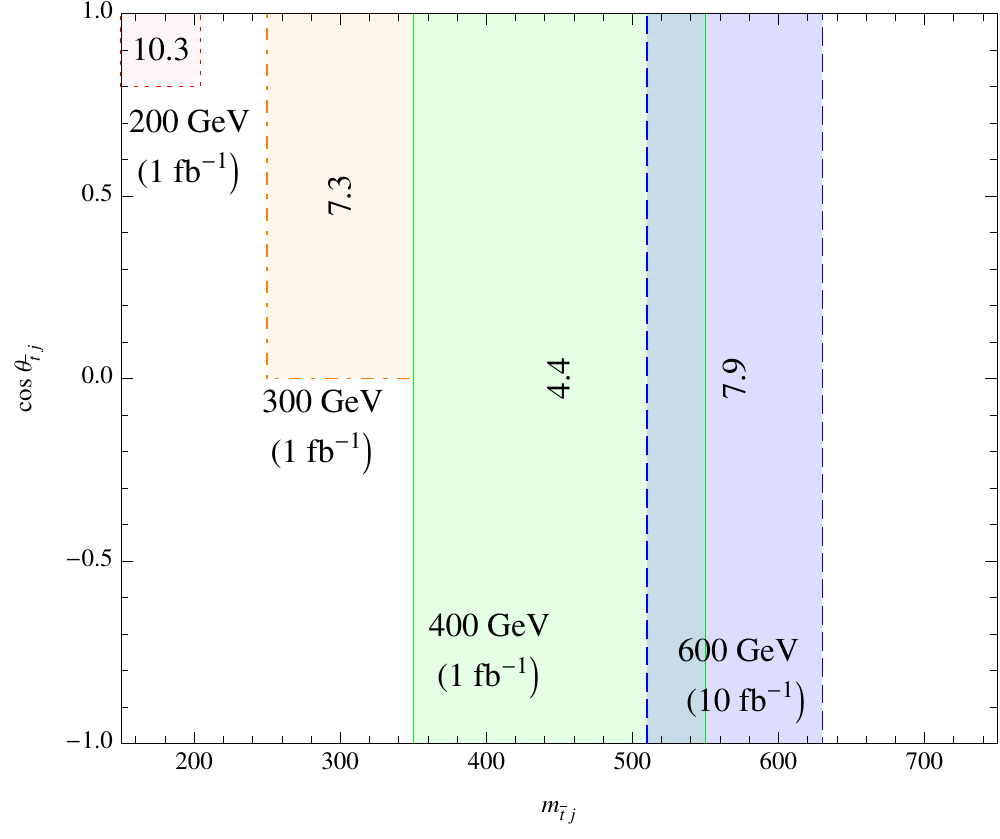} \hspace{2 cm}
\includegraphics[width = 0.4 \textwidth]{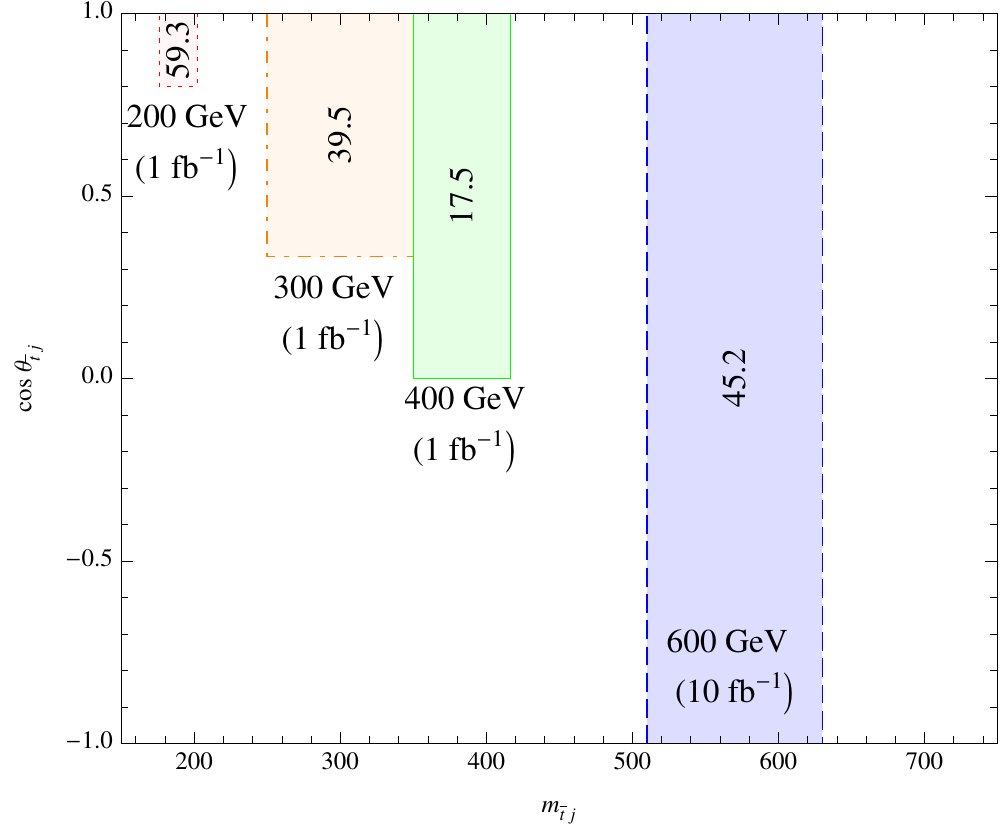}
\caption{ Optimal bins for $W'$ (left) and $Z'_H$ (right) models. The bins shown yielded the largest $\chi^2$ for the background hypothesis given $1~\text{fb}^{-1}$  (for 200, 300, 400 GeV resonances) or $10~\text{fb}^{-1}$  (for a 600 GeV resonance) of data, and coupling $g_R = 1$. This maximum value of $\chi^2$ is shown in the center of each bin. }
\label{best bins}
\end{figure}

\section{Discussion and outlook}\label{sec:4}
Discovery potential for the initial run of the LHC at 7 TeV is summarized in
Fig.~\ref{reach}.  We can see that with 1 fb$^{-1}$, a resonance coupling to $d$ quarks in the initial state and a production cross-section of $\sigma = 40 $ pb for $m_M = 200$ GeV, changing to $\sigma = 4$ pb for $m_M =$ 600 GeV, can be constrained at the $3\sigma$ level.  For example, this corresponds to a reach in coupling of $W'$ to $d_R\bar{t}_R$ at the level of $g_R = 1 $ for $m_{W'} = 200 \mbox{ GeV}$, weakening to $g_R = 1.75$ for $m_{W'} = 600$ GeV, assuming a $100\%$ branching ratio to top-jet.  

For a resonance coupling to $u$ quarks in the initial state and a production cross-section of $\sigma = 27 $ pb for $m_M = 200$ GeV, changing to $\sigma \approx 3 - 4$ pb for $m_M =$ 600 GeV, can be constrained at the $3\sigma$ level with 1 fb$^{-1}$.  
For example, this corresponds to a reach in coupling for a color triplet at the level of $g_\phi = 0.8$ for $m_{\phi} = 200 \mbox{ GeV}$, weakening to $g_\phi = 1.3$ for $m_{\phi} = 600$ GeV, again assuming a $100\%$ branching ratio to top-jet.

\begin{figure}
{\centering
\subfigure[$~W'~$]{\includegraphics[width=0.45\textwidth]{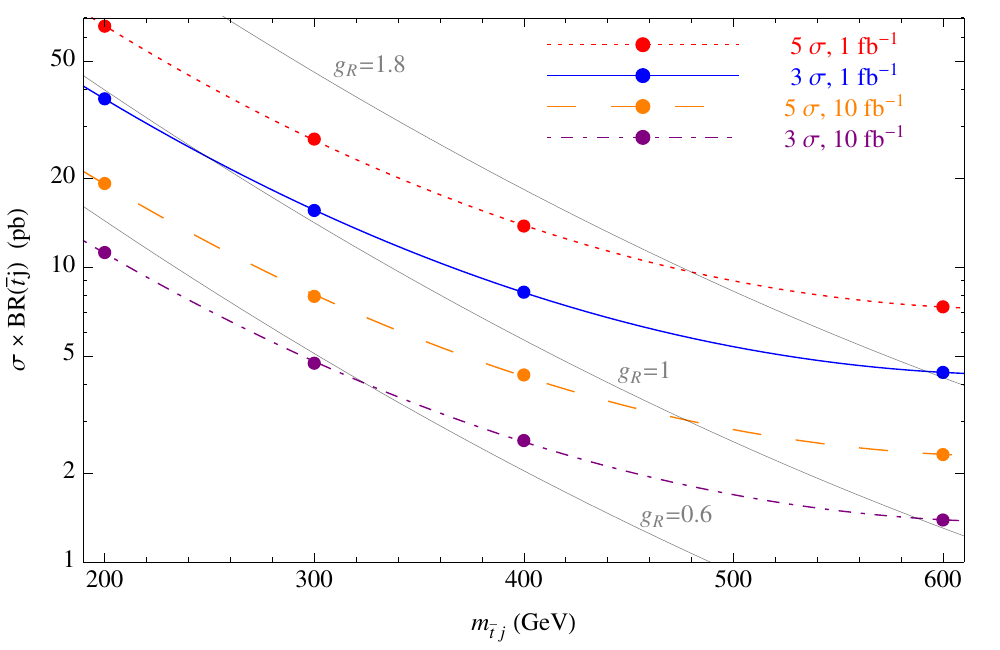}}\qquad
\subfigure[$~Z'_H~$]{\includegraphics[width=0.45\textwidth]{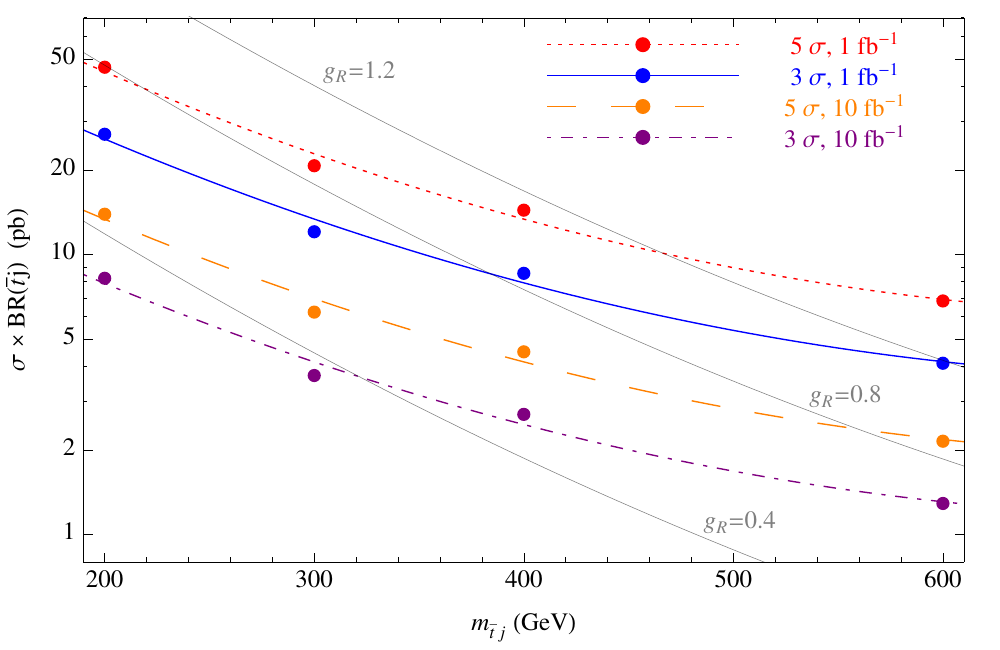}}\\
\subfigure[~Triplet]{\includegraphics[width=0.45\textwidth]{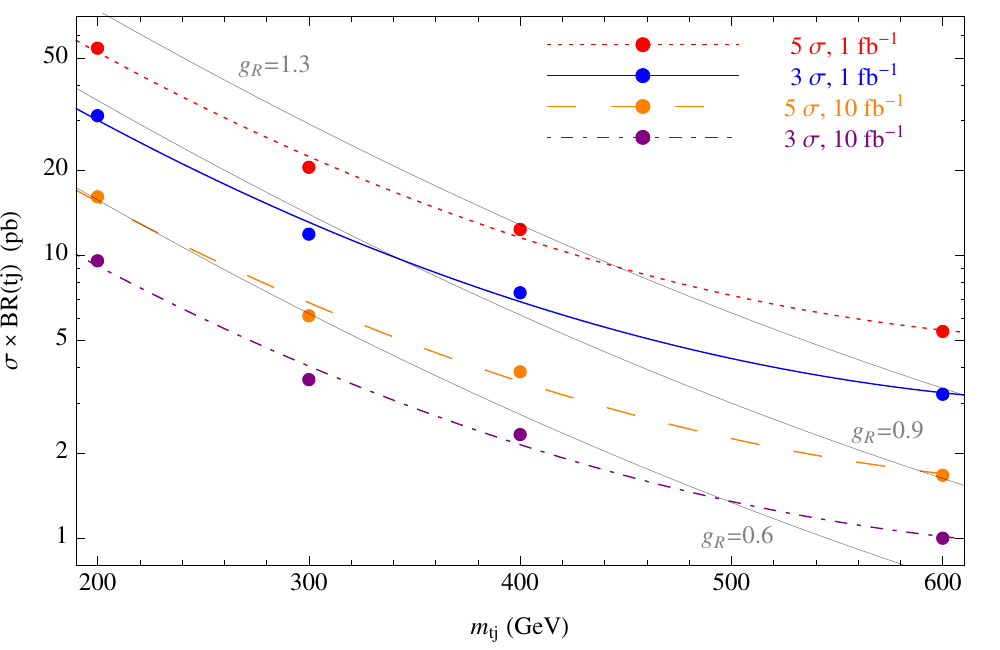}}
\caption{Reach at the 7 TeV LHC for a $W'$ resonance (a), which couples primarily to  down-top, and for a $Z'_H$ resonance (b) and triplet resonance (c), which couple primarily to up-top. Lines of constant coupling $g_R$ as defined in \eqref{lagrangian} are shown in gray, assuming 100\% branching ratios to top-jet. Note that the $W'$ and $Z'_H$ couplings to $\bar{t}_R q_R$ are defined with a factor of $1 / \sqrt{2}$. 
}
\label{reach}
}
\end{figure}

While the purpose of this paper was not to constrain specific models that might generate the Tevatron forward backward asymmetry, we point the reader to existing models in the literature \cite{Cao:2009uz,Cao:2010zb,Choudhury:2010cd,Jung:2009pi,Jung:2010ri,Bai:2011ed} and note that the mass ranges and couplings that were discussed in those papers as sources of the Tevatron forward-backward asymmetry are within reach of the LHC even at 7 TeV and with only 1 fb$^{-1}$ of data.  We leave a detailed discussion of this question in light of recent results from CDF \cite{Aaltonen:2011kc} for a forthcoming paper \cite{citeus}.

\section{Conclusions}\label{sec:5}

We considered the reach of the early LHC operation for top-flavor violating resonances. 
As summarized in Fig.~\ref{reach}, low to moderate mass ($\sim$200--600 GeV) top flavor violating resonances can be effectively identified at the LHC at 7 TeV with only 1 fb$^{-1}$ of data. We constructed a systematic procedure for this task.  Besides searching for resonances in top-jet pairs in $t\bar{t} j$ events, the key feature of this search is a fit to the continuum $t\bar{t}j$ background at moderate to high invariant mass, with a search for bump features in the two-dimensional ${d^2 \sigma \over d m_{\tilde{t} j} d \cos \theta_{\tilde{t} j}}$ distribution at low to moderate invariant mass. Such an early LHC search could shed light on the physics responsible for the top forward-backward asymmetry observed at the Tevatron.

\acknowledgments
We thank Dan Amidei, David Krohn, and Frank Petriello for discussions.

\appendix

\section{Top Quark Reconstruction}\label{top reconstruction}

To find the $tj$ resonance, it is important to identify top and anti-top pairs out of the multi-jet signature with missing $E_T$. In the semi-leptonic decay mode of top quark pairs, top quark momentum as well as neutrino momentum can be fully reconstructed since the missing momentum components are over-constrained by the following on-shell mass relations:
\begin{eqnarray}
y_1 &=& p_\nu^2  = 0,  \label{y1}\\
y_2 &=& (p_\ell + p_\nu )^2 - m_W^2 = 0, \label{y2} \\
y_3 &=& (p_{b_\ell} + p_\ell + p_\nu ) ^ 2 - m_t ^ 2 = 0, \label{y3}\\
y_4 &=& (p_{j_1} + p_{j_2} )^2 - m_W^2 = 0,  \label{y4}\\
y_5 &=& (p_{j_1} + p_{j_2} + p_{b_h})^2  - m_t^2 = 0, \label{y5} 
\end{eqnarray} 
where $\nu$, $\ell$ and $b_\ell$ denote the neutrino, the lepton and the $b$-quark in the leptonic top decay, respectively, and $j_1$,  $j_2$ and $b_h$ denote two jets from $W$ decay and the $b$-quark in the hadronic top decay, respectively. The 4-momentum of $i$th particle is denoted $p_i$.

Since this system is overconstrained, we can find the neutrino momentum that best satisfies the on-shell relations (Eqs.~\eqref{y1}-\eqref{y5}). We use the $\chi^2$ statistic assuming the probabilistic distributions approximately obey Gaussian statistics. The general procedure has been reviewed in the Appendix of \cite{Cheng:2010yy}. For the semi-leptonic top quark pair system, we explicitly show the definition of $\chi^2_{t \bar{t}}$ in Appendix \ref{Sec:Appendix 1}.

Due to the high multiplicity of jets and imperfect $b$-tagging efficiency, the $t\bar{t}j$ system has combinatoric ambiguities when we assign one jet to the leptonic top and three jets among the other jets to the hadronic top, and another remaining jet to the jet from $M \to \tilde{t}j$ decay. We take the five hardest jets including $b$-jets from the event, and consider all combinatoric possibilities.   
Therefore, the reconstruction procedure is to minimize $\chi^2_{t \bar{t}}$ over continuous missing neutrino momentum and the discrete combinatoric possibilities. 
We use the simulated annealing method for minimization. Due to experimental errors, especially in jet momentum measurements and missing transverse energy measurements, this reconstruction often does not lead to the correct combination and thus fails to reconstruct the top momentum. To get a high quality kinematic distribution, we further require a cut on the resultant $\chi^2_{t \bar{t}}$ value. The cut value for $\chi^2_{t \bar{t}}$ we use is 3, but the top reconstruction efficiency is not highly sensitive to this value; we use this value because it corresponds roughly to a 1 $\sigma$ deviation from a perfect fit. The reconstruction efficiencies for four benchmark cases are shown in Table \ref{cross-sections}.

\section{$\chi^2$ statistic in Semi-leptonic Top Pair System} \label{Sec:Appendix 1}

In the semi-leptonic top decays, all momenta except the neutrino momentum are directly measured in the detector. For the neutrino, the transverse directional components are determined by the missing transverse momentum. The longitudinal and time component must be determined as those giving the best fit value of $\chi^2_{t \bar{t}}$ for Eqs.~(\ref{y1})--(\ref{y5}). In this section, we summarize the definition of $\chi^2_{t \bar{t}}$ in a semi-leptonic top pair system.

The $\chi^2_{t \bar{t}}$ statistic represents the likelihood of the hypotheses, Eqs. (\ref{y1})--(\ref{y5}). It is written as 
\begin{eqnarray}
\chi^2_{t \bar{t}} = y^T \cdot V^{-1} \cdot y, 
\end{eqnarray}
where $V_{ij}$ is the covariance matrix between $y_i$ and $y_j$, 
\begin{eqnarray}
V_{ij} = \langle y_i y_j \rangle.  \label{covariance}
\end{eqnarray}
Here, $\langle q \rangle$ for a quantity $q$ is the statistical ensemble average of $q$. Since the equations are generic functions of the measurement variables and the parameters that we will determine from the minimization of $\chi^2_{t \bar{t}}$, we can expand the covariance matrix $V$ to the more detailed form: 
\begin{eqnarray}
 V_{ij} = \frac{\partial y_i}{\partial p_k} \langle p_k p_l \rangle
\frac{\partial y_j}{\partial p_l},  
\end{eqnarray}
where the $p_k$s are momentum and mass parameters. In this case, we have 24 measurement momentum and mass parameters:\footnote{Again, we do not consider the neutrino mass as a probabilistic parameter here.}
\begin{eqnarray}
\{ p^\mu_{b}, p^\mu_{\ell}, p_{j_1}, p_{j_2}, p_{j_3}, \misspt_x \misspt_y, m_W,
 m_t  \},
\end{eqnarray}
where $\misspt_x$ and $\misspt_y$ are the $x-$ and $y-$components of the missing transverse momentum. For massless particles, four components in $p^\mu_i$ of a visible particle $i$ are not independent of each other since we measure particle momentum from the pseudo-rapidity $\eta_i$,  the azimuthal angle $\phi_i$ and the transverse momentum $E_{Ti}$ of each particle $i$ from the tracking chamber and the calorimeter in the detector. Regarding  $\eta_i$, $\phi_i$ and $E_{Ti}$ of each particle as true independent variables, we have 
\begin{eqnarray}
\langle p^\mu_i p^\nu_j \rangle = \delta_{ij} 
\left(\frac{\partial p^\mu_k}{\partial q_I} \right)\cdot
\left( 
\begin{array}{ccc}
\delta \eta_i^2 &  & \\
&  \delta \phi_i^2  & \\
 &  & \delta E_{Ti}^2
\end{array}
\right) \cdot \left(\frac{\partial p^\nu_l}{\partial q_J}\right),  \label{oneparticlecov}
\end{eqnarray} 
where $q_I$ is a collective notation for $\eta_i$, $\phi_i$ and $E_{Ti}$ and $\delta \eta_i$, $\delta \phi_i$ and $\delta E_{Ti}$ denote the standard deviation of each observable, respectively. 
We summarize estimated errors of visible identified objects in the LHC detectors (especially for the CMS detector) in Appendix \ref{CMS}. 

The missing transverse momentum depends on other momentum observables
since 
\begin{eqnarray}
\vec{\misspt} = - \sum_{i} \vec{p}_{Ti}, 
\end{eqnarray}
by definition, where the index $i$ runs over all of the visible particles. The covariance matrix related to $\misspt$ is
\begin{eqnarray}
\langle \vec{\misspt}_{x,y} \vec{p}_{i\,x,y} \rangle &=& 
- \langle \vec{p}_{i\,x,y} \vec{p}_{i\,x,y} \rangle \\
\langle \vec{\misspt}_{x,y} \vec{\misspt}_{x,y} \rangle &=& 
\sum_i \langle \vec{p}_{i\,x,y} \vec{p}_{i\,x,y} \rangle, 
\end{eqnarray}
where $\sum_i$ denotes the summation over all of the visible particles.

Therefore, the $24 \times 24$ covariance matrix in Eq. (\ref{covariance}) is  of the form 
\begin{eqnarray}
&& \langle p_k p_l  \rangle = \\
&&
\left(
\begin{array}{ccccccccc}
\langle p_b p_b \rangle & & & & &  \langle p_b \misspt \rangle  & & & \\
     & \langle p_\ell p_\ell \rangle & & & & \langle p_\ell \misspt \rangle & & &  \\
 & & \langle p_{j_1} p_{j_1} \rangle & & & \langle p_{j_1} \misspt \rangle  & & & \\ 
 & & & \langle p_{j_2} p_{j_2} \rangle & & \langle p_{j_2} \misspt \rangle & & & \\
 & & &  & \langle p_{j_3} p_{j_3} \rangle & \langle p_{j_3} \misspt \rangle &
 & &  \\ 
 \langle p_b \misspt \rangle & \langle p_\ell \misspt \rangle & \langle p_{j_1} \misspt \rangle & \langle p_{j_2} \misspt \rangle & \langle p_{j_3} \misspt \rangle & \langle \misspt \misspt \rangle & & \\
 & & & & & & & \delta m_W^2  & \\
 & & & & & & &               & \delta m_t^2   
\end{array}
\right)
\end{eqnarray}

Now that we have defined $\chi^2_{t \bar{t}}$, we need to find the configuration that gives rise to the minimum. We have two unknown parameters $p^0_\nu$ and $p^3_\nu$ from the neutrino. By minimizing $\chi^2_{t \bar{t}}$, we determine these unknown parameters. The degrees of freedom in this system are thus $5-2 = 3$.

\section{Experimental Error from CMS TDR} \label{CMS} 

 The experimental errors from the measurements of particle momenta and missing energy are summarized in Table \ref{app:detector}. 
\begin{table}
\centering
\begin{tabular}{|cc|}
\hline
\multicolumn{2}{|l|}{Electrons and muons :}\\
\hline
 & $|\eta|< 2.5,\ \ p_T>10,$ \\
 & $\frac{\delta p_T}{p_T}=0.008\oplus 0.00015\,p_T,$ \\
 &  $\delta\theta = 0.001, \delta\phi = 0.001.$  \\
\hline
\multicolumn{2}{|l|}{Jets: } \\
\hline
&$\frac{\delta_{E_T}}{E_T}=\left\{
    \begin{array}{cl}\frac{5.6}{E_T}\oplus\frac{1.25}{\sqrt{E_T}}\oplus0.033, &\mbox{for }|\eta|<1.4,\\
                      \frac{4.8}{E_T}\oplus\frac{0.89}{\sqrt{E_T}}\oplus0.043, &\mbox{for }1.4<|\eta|<3.0, \\
                     \end{array}\right.$\\
&$\delta\eta=0.03, \ \ \delta\phi=0.02 \ \ \mbox{for }|\eta|<1.4,$\\
&$\delta\eta=0.02, \ \ \delta\phi=0.01\ \ \mbox{for }1.4<|\eta|<3.0.$\\
\hline
\end{tabular}
\caption{ 
Errors from the measurements of particle momenta used in our analysis. The observables of energy dimension are in GeV units and the angular and the rapidity variables are in radians. Here,  $A \oplus B \equiv \sqrt{A^2 + B^2}$. For jets, errors are taken from the CMS TDR \cite{Ball:2007zza}. For electrons and muons, the resolution here corresponds roughly to the CMS tracking system performance in the central region ($\eta=0$). The resolution becomes slightly worse at higher rapidity until $|\eta|\gtrsim2$ where it starts to diverge \cite{Ragusa:2007zz}. We ignore this rapidity dependent effect. For photons, the resolution for position measurement corresponds to the CMS ECAL performance obtained using electron beams ($10<p_T<50\GeV$). }
\label{app:detector}
\end{table}

\end{document}